\documentclass[%
aip,
amsmath,amssymb,
reprint,%
twocolumn,
]{revtex4-2}

\usepackage{graphicx}
\usepackage{dcolumn}
\usepackage{bm}
\usepackage{float}
\usepackage{subfigure}
\usepackage{color,soul}
\usepackage{gensymb}
\usepackage{lipsum}
\usepackage{amsmath}
\usepackage{amsfonts}
\usepackage{lineno}
\usepackage{cancel}
\usepackage[normalem]{ulem}
\usepackage[section]{placeins}

\begin{document}
\preprint{}

\title{Impact of ion-beam stopping power on proton-boron fusion yield in the pitcher-catcher scheme driven by ultra-intense laser}

\author{J. Y. Hua}
\affiliation{State Key Laboratory of Ultra-intense laser Science and Technology, Shanghai Institute of Optics and Fine Mechanics, Chinese Academy of Sciences, Shanghai 201800, China}
	
\author{X. F. Li}\email{xiaofengli@siom.ac.cn}%
\affiliation{State Key Laboratory of Ultra-intense laser Science and Technology, Shanghai Institute of Optics and Fine Mechanics, Chinese Academy of Sciences, Shanghai 201800, China}
\affiliation{University of Chinese Academy of Sciences, Beijing 100049, China}
	
\author{J. X. Wang}
\affiliation{School of Physics, East China Normal University, Shanghai 200241, China}

\author{Y. X. Leng}
\affiliation{State Key Laboratory of Ultra-intense laser Science and Technology, Shanghai Institute of Optics and Fine Mechanics, Chinese Academy of Sciences, Shanghai 201800, China}
\affiliation{University of Chinese Academy of Sciences, Beijing 100049, China}
	
\author{Y. Tian}\email{tianye@siom.ac.cn}%
\affiliation{State Key Laboratory of Ultra-intense laser Science and Technology, Shanghai Institute of Optics and Fine Mechanics, Chinese Academy of Sciences, Shanghai 201800, China}
\affiliation{University of Chinese Academy of Sciences, Beijing 100049, China}
	
\author{R. X. Li}
\affiliation{State Key Laboratory of Ultra-intense laser Science and Technology, Shanghai Institute of Optics and Fine Mechanics, Chinese Academy of Sciences, Shanghai 201800, China}
\affiliation{University of Chinese Academy of Sciences, Beijing 100049, China}
\affiliation{School of Physical Science and Technology, ShanghaiTech University, Shanghai 201210, China}

\date{\today}
	
\begin{abstract}
	The effect of stopping power on proton-boron fusion is investigated for a proton beam propagating through boron plasma. Due to the stopping power, the electron temperature of the boron target rises as the proton beam deposits energy. Consequently, a feedback mechanism becomes significant when the trailing part of the beam propagates into the preheated plasma. By coupling this phenomenon with the proton-boron fusion process, fusion yields are systematically investigated by varying the central energy of the proton beam, as well as the thickness and density of the boron target. It is found that, under the influence of stopping power, the optimal central energy for fusion deviates from the intrinsic 672 keV resonance and shifts to approximately 900 keV.  Moreover, the present results are substantiated by particle-in-cell simulations, which provides a valuable reference for subsequent high‑yield hydrogen‑boron fusion.
\end{abstract}

\maketitle

\section{INTRODUCTION}
Controlled nuclear fusion is widely regarded as the ultimate solution for the energy crisis, as it offers abundant fuel, clean energy, and high energy density. The recent achievement of net energy gain at the NIF facility, alongside the continuous progress in Tokamak research, has further validated the scientific feasibility of controlled fusion\cite{xu2023experimental,gibney2022reactor,zylstra2021record,gus2021extreme,liu2021fabrication,tikhonchuk2021studies,zylstra2022burning,zylstra2022experimental,weber2021recent,atzeni2022breakthrough}. It marks a major milestone toward the fusion era, laying a solid foundation for future clean energy applications\cite{betti2023milestone,hurricane2023physics}. Besides D-T(Deuterium-Tritium) fusion, p-$^{11}$B (proton-boron) fusion has garnered significant attention due to its aneutronic nature\cite{hora2021elimination,becker1987low,segel1965states,nevins2000thermonuclear,stave2011understanding,moreau1977potentiality}, where it involves a proton colliding with a boron nucleus to produce three alpha particles. This process is considered as a highly promising candidate for power generation, since all the reaction products (only alpha) are charged particles. Moreover, owing to the substantial terrestrial abundance of both proton and boron, this fusion scheme offers distinct strategic advantages over D-T scheme.
However, the p‑¹¹B fusion cross‑section peaks only at plasma temperatures of several hundred keV, a regime where bremsstrahlung losses become exceedingly severe, thus compromising the energy balance and making the reaction not self-sustaining under thermal equilibrium\cite{xie2024bremsstrahlung,munirov2023suppression}.

Non‑thermal‑equilibrium fusion schemes have been actively pursued to circumvent this intrinsic bottleneck, particularly in the context of intense laser–plasma interactions\cite{hu2025nonequilibrium}. This regime, characterized by particle temperatures deviating from the Maxwellian distribution, is currently the most widely pursued for igniting p-$^{11}$B fusion. Based on target configuration, these non-thermal approaches are generally categorized into two types. The first, known as the ``in-target" scheme, employs a solid target containing both protons and boron\cite{bonvalet2021energetic,hegelich2023photon}. Upon laser irradiation, the target is heated to form a localized hot plasma, initiating proton-boron fusion reactions. The second, termed the ``pitcher-catcher" scheme, utilizes a dual-component target typically consisting of a plastic or proton-rich target and a separate boron target. In this configuration, laser irradiation of the plastic ``pitcher" accelerates protons, these protons then traverse the inter-target space and induce fusion reactions upon impacting the boron ``catcher-target"\cite{belyaev2005observation,labaune2013fusion,picciotto2014boron}. 

Over the past decades, the rapid maturation of high‑power laser technology has progressively made p-$^{11}$B nuclear reactions experimentally feasible.
As demonstrated by Belyaev, \textit{et al.}\cite{belyaev2005observation},  a yield of $10^4$
alpha particles per joule was obtained from boron-rich polyethylene planar targets irradiated with the Nd:glass laser system.  More recently, experimental investigations at the LFEX facility have achieved sig nificantly higher yields in the range of $10^6-10^7 \alpha/\mathrm{J}$\cite{bonvalet2021energetic,margarone2022target}. However, the detected alpha particle yields in experiment remain significantly below the critical gain threshold of $2 \times 10^{12}\alpha/\mathrm{J}$, indicating an urgent need for further investigation\cite{belyaev2005observation,bonvalet2021energetic,margarone2022target,giuffrida2020high,picciotto2014boron,labaune2013fusion}. In particular, numerous challenges still need to be addressed for the aforementioned non-thermal equilibrium fusion scheme, such as the design configurations of solid targets system \cite{wang2025proton,liu2023enhancement}, and the underlying physics of the beam-target interactions, including stopping power and energy-loss dynamics\cite{zhang2022ion,gericke2002stopping}.

The stopping power, originally conceived for radiation protection, has become a pivotal quantity in high‑energy‑density physics, where it is now critical for assessing beam‑energy deposition and the resulting yield limitation in beam–target dominated fusion schemes. Over the past few decades, the theory of stopping power of single particles in plasmas and solid targets has been extensively investigated and is now well established\cite{ziegler1999stopping,gus2009method,deutsch2016ion,deutsch2010ion,betz1984ion,bailey1983ab,peter1991energy,barges2025modeling}. Currently, some researchers have incorporated stopping power into the physical models of p-$^{11}$B fusion\cite{mcguire1973procedure,liu2024proton,eliezer2020mitigation}. However, these studies primarily rely on single-particle models to explain experimental phenomena or propose novel fusion schemes, such as grid targets\cite{liu2023enhancement} or electric field acceleration scheme\cite{eliezer2020mitigation} to optimize the proton energy near the maximum of the proton-boron fusion cross-section, which peaks at approximately 672 keV.  The studies on the collective effects of stopping power induced by ion beams remains insufficient \cite{gericke2002stopping,gericke1999beam,deutsch1984atomic}. The majority of previous investigations have addressed proton transport in plasma  without considering the feedback effect of the proton energy on the plasma. Moreover, a self-consistent coupling between the beam stopping power and the fusion reaction dynamics remains elusive in current theoretical frameworks.

This work introduces an advanced model for ion stopping with feedback effect in fully ionized plasma targets. By unraveling the functional link between the energy deposition profile (and hence the fusion rate) and the incident proton energy, we precisely pinpoint the optimal initial energies for yield optimization.
The validity of this theoretical model, together with the proposed target configuration and laser parameters, is substantiated by Particle-in-Cell (PIC) simulations. Through extensive parameter optimization, we achieved highly efficient ion acceleration driven by laser\cite{xu2017plasma}, generating a proton beam peaked at around 900 keV. Subsequently, the protons undergo deceleration within the boron target, yielding a higher fusion yield than that of the 672 keV incident proton beam.
The paper is organized as follows: Section I presents the background and motivation for this study. Section II details the physical model of stopping power for proton beam. Section III discusses the PIC result with its underlying physics. Section IV provides a summary of our findings and conclusions.

\section{Theoretical model of Stopping Power for a proton beam}
 The stopping power plays a critical role in determining the efficiency of the  p-$^{11}$B fusion reaction, especially within the pitcher-catcher target. As the laser-accelerated protons propagate into the high-density boron slab, their kinetic energy is dissipated primarily through collisions with plasma electrons, alongside significant energy loss to the boron ions via nuclear stopping.
Generally, the stopping force experienced by a projectile (\textit{i.e.}, the incident proton) as it traverses a medium can be attributed to two distinct components: contributions from bound electrons and contributions from free electrons within the plasma. For the bound-electron contribution, it is primarily described by the Bethe-Bloch equation\cite{ziegler1999stopping,gus2009method,deutsch2016ion,deutsch2010ion,betz1984ion,bailey1983ab,peter1991energy}.

In the pitcher-catcher scheme, the intense laser pulse not only accelerates high-energy protons but also generates a concomitant population of hot electrons from the pitcher target. These hot electrons, arriving at the catcher target ahead of the proton beam, pre-heat the target into a dense, highly ionized plasma state. Consequently, the stopping of the subsequently arriving proton beam is largely determined by its interaction with the free electrons in the plasma, which can be described as\cite{mehlhorn1981finite}
\begin{equation}\label{eq4}
	\begin{split}
    \left( \frac{\mathrm{d}E}{\mathrm{d}x} \right)_{free} = \frac{\omega_p^2 Z_{1eff}^2 e^2}{c^2 \beta^2} G(y_e) \ln \Lambda_{free}, \end{split}
\end{equation}
where $\beta = v/c$ is the normalized particle velocity and $n_0$ is the plasma density. The function $G(y_e)$ is given by $\mathrm{erf}(\sqrt{y_e}) - 2\sqrt{y_e/\pi}\exp(-y_e)$, and the parameter $y_e = m_e c^2 \beta^2 / (2T_e)$, where $T_e$ is the plasma temperature. In addition, the plasma frequency is $\omega_p^2 = 4\pi n_0 e^2 Z_2 / (m_e A_2)$, and the free collision parameter is $\Lambda_{\text{free}} = 0.764 \beta c / (b_{\min} \omega_p)$. The effective charge $Z_{1\text{eff}}$ is calculated via $Z_1 [1 - 1.034 \exp(-137.04\beta/Z_1^{0.69})]$, and $b_{\min} = \max[2Z_1/(m_{12}v^2), \hbar/(2m_{12}v)]$ is the minimum impact parameter, where $Z_1$ is the atomic number,  $m_{12}$ is the reduced mass, $Z_2$ and $A_2$ correspond to the charge and nucleon numbers of the plasma ion.

In this work, we investigate the stopping power of a proton beam propagating  along the positive x-axis through a boron target. The phase-space distribution function of the proton beam can be denoted as $f_p(x, v, t)$, and its dynamics was described by the one-dimensional Fokker-Planck equation\cite{risken1989fokker}:
\begin{equation}\label{eq4}
\frac{\partial f_p}{\partial t} + v \frac{\partial f_p}{\partial x} = - \frac{\partial}{\partial v} \left[ \left( \frac{\mathrm{d}E}{\mathrm{d}x} \right)_{free} f_p \right].
\end{equation}

Owing to the conservation of energy, the energy lost by the incident protons are transferred to the electrons. Furthermore, since the velocity of the electrons after the collision is much lower than that of the incident ions\cite{du2026effects}, their temperature can be approximated as:
\begin{equation}\label{eq4}
\frac{\partial T_e(x, t)}{\partial t} = \frac{1}{n_0} \int_{0}^{\infty} \left( \frac{\mathrm{d}E}{\mathrm{d}x} \right)_{free} m_p v f_p(x, v, t) \mathrm{d}v.
\end{equation}

The stopping power of the proton beam in the plasma, along with its heating effect on plasma electrons, is calculated by using Eqs. (2) and (3).  The resulting distribution function $f_p(x, v, t)$ can provide a basis for assessing  the fusion yield. The detailed procedure is as follows,
\begin{equation}\label{eq4}
Y = \iiint f_p(x, v, t) n_0 \langle \sigma v \rangle \mathrm{d}x \mathrm{d}v \mathrm{d}t,
\end{equation}
where $\sigma$ represents the p-$^{11}$B reaction cross-section \cite{nevins2000thermonuclear}, and $v$ is the proton velocity derived from its kinetic energy ($E$), which decreases according to the stopping power $\mathrm{d}E/\mathrm{d}x$. The boron ions within the target are assumed to be at rest. Following the calculation of the fusion yield, the corresponding fusion gain can be evaluated as:
\begin{equation}\label{eq4}
G = Y Q / E_{\text{total}},
\end{equation}
where {$Q$}=8.7 MeV is the reaction energy yield per proton-boron event, and {$E_{\text{total}}$} is the total energy of the inject proton beam.

\begin{figure}[t]
	\includegraphics[width=0.48\textwidth]{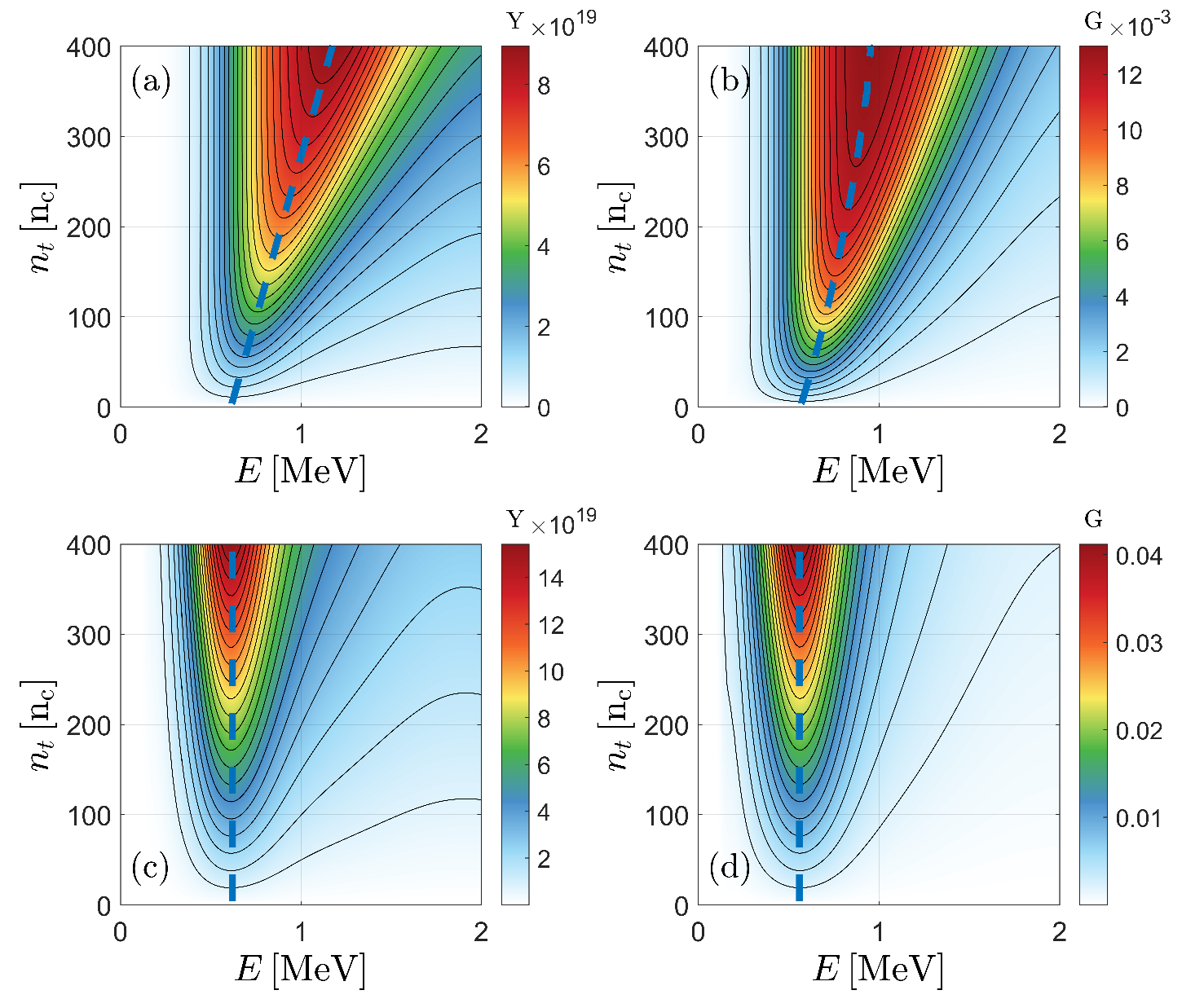}
	\caption{Dependence of fusion yield $Y$ on proton beam center energy for boron targets with varying density but constant thickness ($L=100$ $\mu\text{m}$): (a) including stopping power effects and (c) neglecting them. While (b) and (d) show the fusion energy gain $G$ corresponding to the cases in (a) and (c), respectively.}\label{fig1}
\end{figure}

In the present work, we assume the proton beam density follows a  Gaussian spatial profile $n_b \exp\left[-(x - x_0)^2 / x_R^2\right]$, where the centroid position $x_0 = -12~\mu\text{m}$, $x_R = 3~\mu\text{m}$, and the peak density $n_b = 12 n_c$ (with $n_c = 1.74 \times 10^{27}~\text{m}^{-3}$).  The energy spectrum of the proton beam also follows a Gaussian distribution $\propto \exp\left[-(E - E_0)^2 / \Delta E^2\right]$, characterized by a central energy $E_0$ and an energy spread of $\Delta E / E_0 = 25\%$. The boron target is set with a uniform density of $n_t$ and spans the spatial region from $x = 0$ to $L$, where $L$ is the target thickness. 
To evaluate the overall fusion yield throughout the process, we firstly present the fusion yield of protons traversing the target by varying its density or thickness. As illustrated in Fig. 1, the fusion yield $Y$ and energy gain $G$ exhibit distinct dependencies on the proton central energy and target density with a fixed target thickness of $L=100$ $\mu\text{m}$.  For any given target density, the fusion yield as a function of proton beam energy is non-monotonic, with a well-defined optimum that maximizes the yield. Specifically, Fig. 1(a) demonstrates that this optimal center energy increases continuously with rising target density. Notably, the peak position of the maximum fusion yield $Y$ (the blue dashed line) does not remain fixed near 672 keV; instead, it shifts gradually to higher energies as the boron density increases. Furthermore, the corresponding energy gain $G$ is shown in Fig. 1(b). To maintain a high gain value, the optimal proton energy tends to converge towards a value over 900 KeV at higher densities. 

\begin{figure}[t]
	\includegraphics[width=0.48\textwidth]{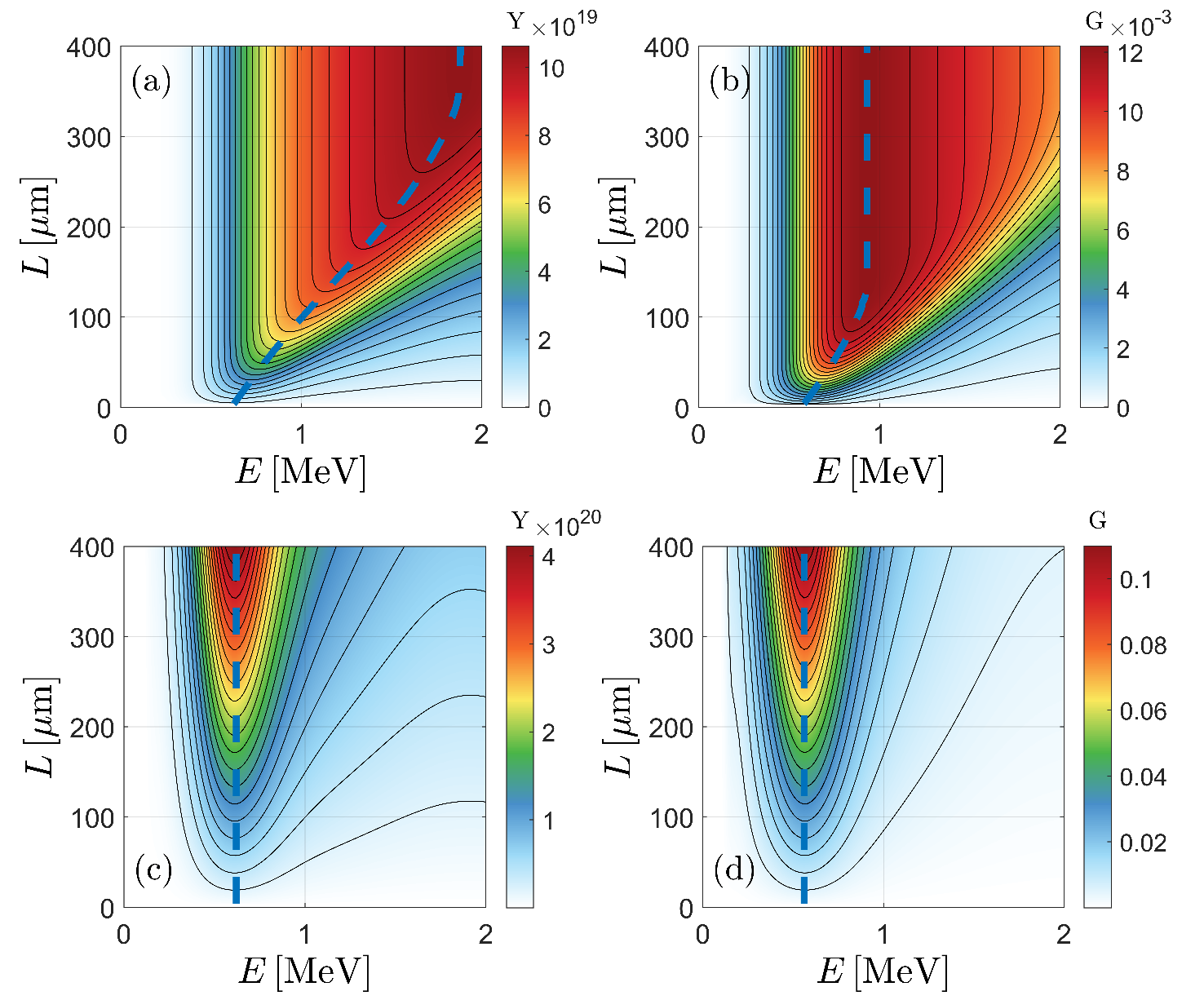}
	\caption{Fusion yield $Y$ as a function of proton beam center energy for targets with varying thickness but constant density ($n_b=120n_c$): (a) with stopping power and (c) without stopping power. While (b) and (d) show the fusion energy gain $G$ corresponding to the cases in (a) and (c), respectively.}\label{fig2}
\end{figure}

\begin{figure*}[t]
	\includegraphics[width=0.8\textwidth]{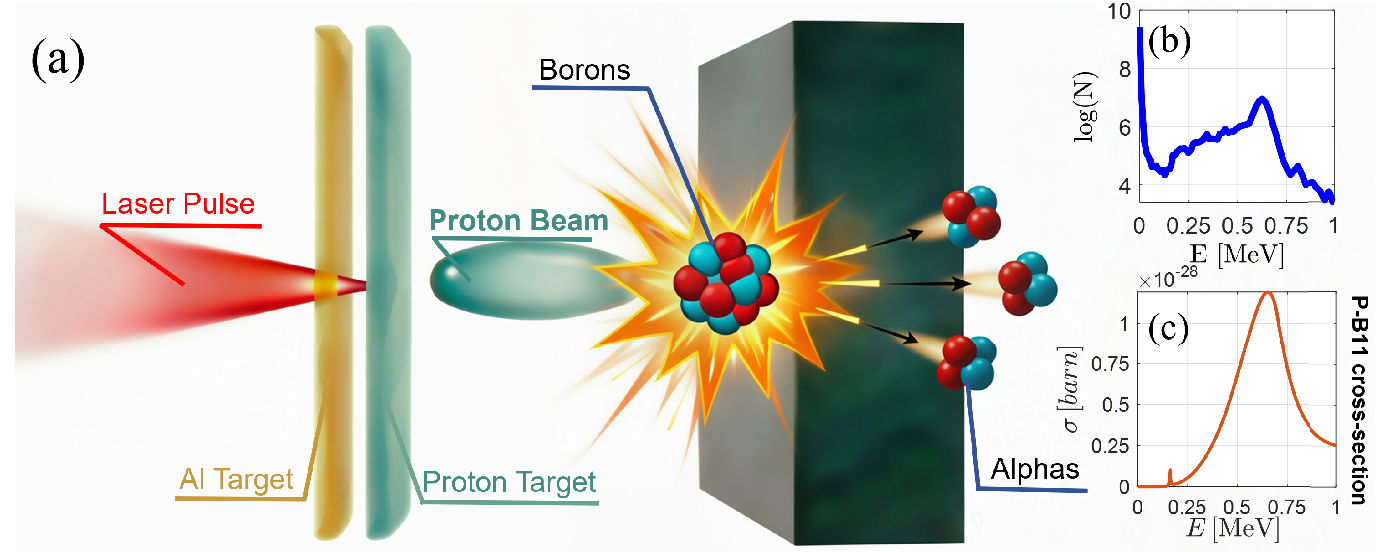}
	\caption{(a) Schematic of the laser-driven pitcher-catcher configuration. A high-intensity laser pulse impinges on the double foil-targets consisting of aluminum and hydrogen layers (i.e, the pitcher target), generating a proton beam. The accelerated protons then travel through the rear side and interact with the boron catcher target.   (b) The energy spectra of the proton beam. (c) The fusion cross section of p-$^{11}$B fusion\cite{nevins2000thermonuclear}.}\label{fig3}
\end{figure*}

These phenomena can be attributed to the interplay between the collective stopping power and the finite target length. In a sufficiently dense target, high-energy protons (\textit{e.g.} $\geq$ 1 MeV) can be effectively decelerated via collisions with target electrons, allowing them to reach the energy corresponding to the peak of the fusion cross-section (672 keV). The maximum fusion yield is achieved under the condition of complete proton energy deposition within the target. This implies that the target density must be carefully optimized; only when the target is sufficiently dense to completely stop the proton beam can the maximum fusion yield be achieved. However, without considering the stopping power, the behavior changes significantly, as illustrated in Figs. 1(c) and (d). Since the proton energy remains constant within the target, the fusion yield depends primarily on the fusion cross-section and the target density which can be simply expressed as $ \propto n_0 \langle \sigma v \rangle \,$. Consequently, the observed characteristics align perfectly with the peak of the p-$^{11}$B reaction cross-section at 672 keV. Furthermore, both the reaction rate and energy gain are significantly higher than them in cases with stopping power.

Furtherly, the effect of target thickness on the fusion yield is investigated by fixing the target density at $n_b=120n_c$, as shown in Figs. 2. As the target thickness $L$ increases, the optimal incident proton energy for the maximum fusion yield gradually shifts to higher values, as revealed in Fig. 2(a), and this trend is similar to that in Fig. 1(a).  The maximum fusion yield is attained when the proton beam deposits its full energy within the target, which requires a target thickness sufficient to completely stop the beam.  Additionally, the energy gain analysis presented in Fig. 2(b) demonstrates that the optimal proton energy converges to approximately 900 keV.   Without considering the stopping-power effect, the proton energy yielding the maximum fusion yield is approximately 672 keV, which is consistent with the results shown in Figs. 1(c) and 1(d).

 These results indicate that, under conditions of sufficiently large target thickness and density, the stopping power significantly influences the beam energy evolution. This effect causes the optimal initial proton energy to deviate from the intrinsic resonance peak of 672 keV toward a higher value, approximately 900 keV. To elucidate these observations, the p-$^{11}$B reaction cross-section is examined, featuring a sharp resonance peak at 672 keV, a precipitous decline in the sub-resonance regime, and a broad supra-resonance plateau at ~20$\%$ of the peak value. In dense targets, a 672 keV proton immediately falls below this sharp resonance peak due to stopping power, severely suppressing the fusion yield. Conversely, protons with initial energies above 672 keV will initially pass through the broad resonance plateau on the cross-section curve as they decelerate, before finally reaching the resonance peak. This extended path through the energy range of appreciable cross section results in a larger integrated effective cross section and thus a higher fusion yield.
 
Notably, this behavior is significantly modified by temperature feedback effects that emerge in the case of intense, long-pulse proton beams.  Firstly, protons at the beam forefront transfer energy to electrons via stopping power, elevating electron temperatures. After that, the elevated electron temperatures reduce the subsequent stopping power experienced by trailing protons. This reduction effectively lowers the average stopping power across the entire beam, enabling the proton ensemble to maintain higher energies for longer. Consequently, they traverse more of the favorable resonance plateau, ultimately enhancing fusion yield.   In contrast, when the target density and thickness are sufficiently low, energy loss effects become negligible. Under such conditions, the fusion yield is dominated by the intrinsic fusion cross-section, with the energy deposition profile playing a minor role.  Accordingly, the maximum yield is achieved as long as the incident proton energy covers the 672 keV resonance peak, without the need for exact resonance matching.

\section{Simulation results \& discussions}

In order to verify the theoretical predictions, a series of  one-dimensional simulations was carried out by the PIC code EPOCH \cite{arber2015contemporary}, employing a spatial high resolution of 600 cells per micrometer and 64 macroparticles per cell, in order to accurately model the electron-proton collision dynamics\cite{arber2015contemporary} and to account for the stopping power effect. The interaction scheme is presented at Fig. 3, where a three-layer target scheme is used. The first target, an Aluminum plasma, was positioned from $x = 5~\mu\text{m}$ to $10~\mu\text{m}$ with a thickness of $5~\mu\text{m}$ with an exponential density ramp from 0 to $1 n_c$. The proton target was located between $x = 11.7~\mu\text{m}$ and $13.5~\mu\text{m}$, possessing a thickness of $1.7~\mu\text{m}$. The third target, a solid boron slab with a density of $160n_c$, was placed from $x = 15~\mu\text{m}$ to $115~\mu\text{m}$. The system was irradiated by a circularly polarized laser pulse with a wavelength of $\lambda = 0.8~\mu\text{m}$, a pulse duration of $16.8T_0$ with a Gaussian distribution, where $T_0$ is the duration of laser cycle. This multi-layer target structure in this scheme effectively boosts the accelerated proton yield, resulting in a 10\% increase over the case without the first target\cite{xu2017plasma}.

 The proton acceleration process resembles a typical radiation pressure acceleration mechanism, described by  \cite{macchi2005laser,macchi2013ion,xu2017plasma} $v_a/c = \sqrt{(Z m_e n_c)/(A m_p n_e)} a_0$, where $v_a$ is the velocity of the ion particles, $Z$ is the ionic charge number, $A$ is the ionic mass number, $m_e$ is the electron mass, $m_p$ is the proton mass, $n_c$ corresponds to the critical density of the laser, and $n_e$ corresponds to the target density. This equation can also be transformed into $E_i = 0.5m_p v^2 = 0.5Z m_e c^2 a_0^2 n_c / (A n_e)$. With a laser with a normalized amplitude of $a_0 = 10.5$, we obtain a proton with a peak energy of 672 keV (Fig. 3(b)),  which agrees well with the resonance peak of the p-$^{11}$B fusion cross-section (Fig. 3(c)).

\begin{figure}[t]
	\includegraphics[width=0.48\textwidth]{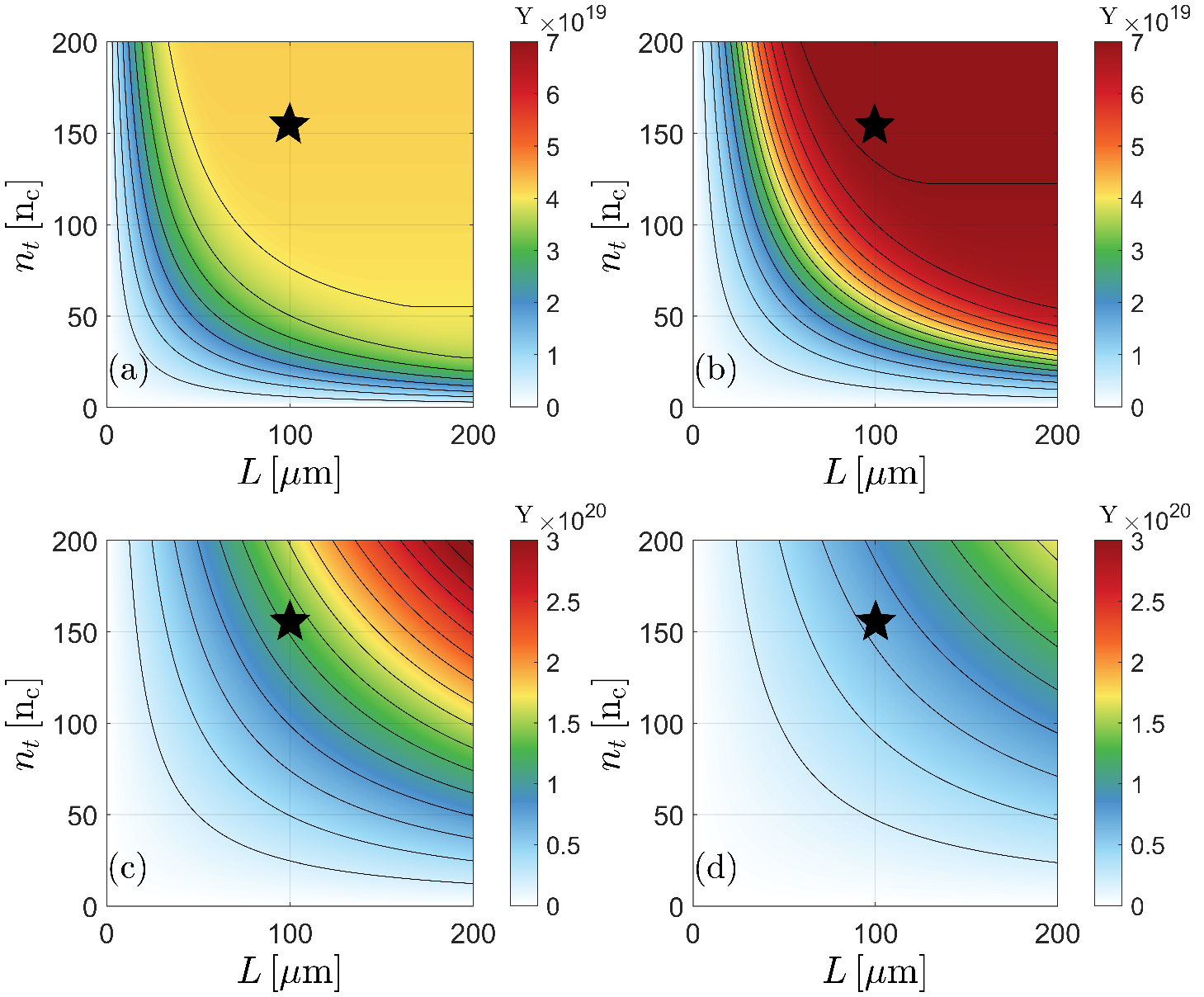}
	\caption{Fusion yield $Y$ versus boron target density and thickness. The stopping power of protons is taken into account in (a) for 672 keV and (b) for 900 keV, whereas (c) and (d) show the counterparts without stopping-power effects. The black star highlights the parameters used for the subsequent simulations.}\label{fig4}
\end{figure}

\begin{figure}[b]
	\includegraphics[width=0.32\textwidth]{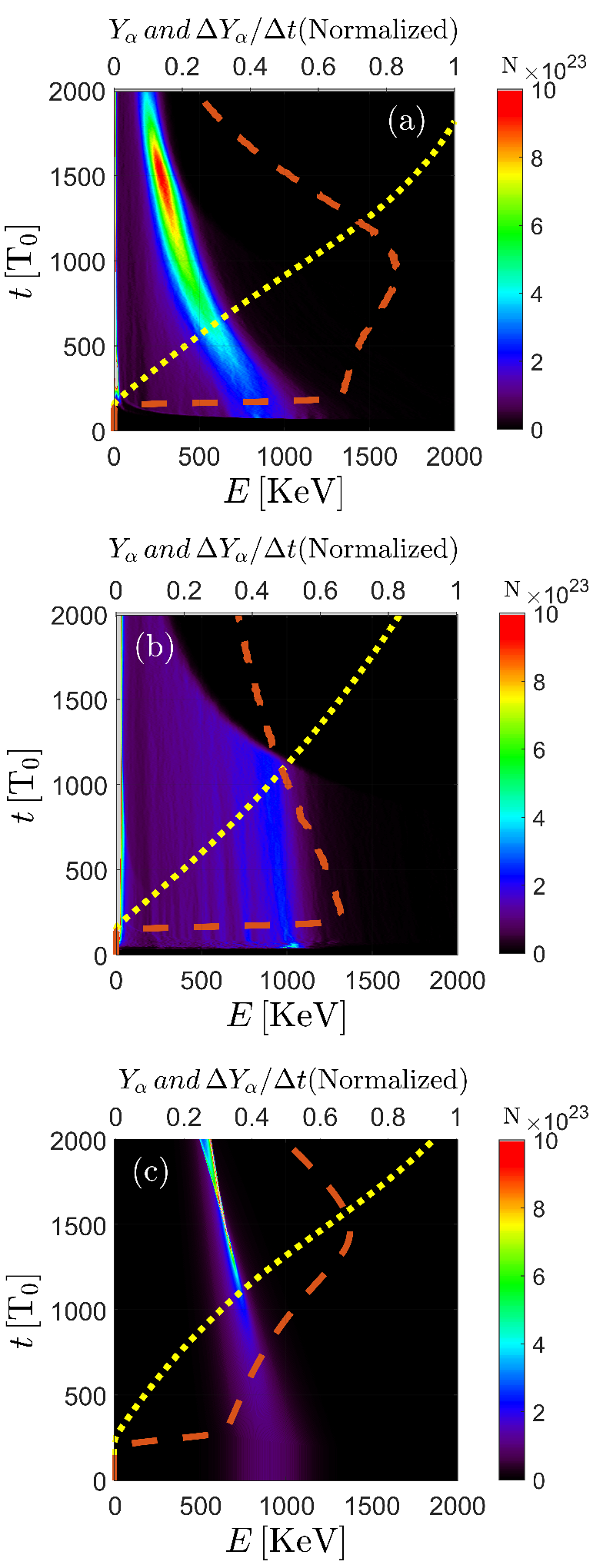}
	\caption{ Comparative analysis of the $\alpha$-particle yield generated by a 900 keV proton beam interacting with a boron target, where the laser intensity $a_0=11.5$. The simulation results in (a) include the stopping power of protons, whereas those in (b) are obtained by switching off this energy-loss mechanism. (c) the theoretical prediction based on Eq. (6). The colormap shows the time-resolved proton energy spectra. The orange and yellow dashed lines denote the cumulative and instantaneous $\alpha$-particle yields, sampled at intervals of $5 T_0$.}
\end{figure}

Based on  the theoretical model detailed in Section II, we firstly evaluate the  fusion yields of fixed-energy proton beams in boron targets as functions of target density and thickness, as presented in Figs. 4. Achieving the maximum fusion yield requires that the protons be fully stopped within the target. Consequently, as anticipated, higher-energy protons demand correspondingly thicker and denser targets to achieve this optimal stopping condition. Notably, when stopping power effect is included, protons with an initial energy of 900 keV produce a higher fusion yield than those at 672 keV, in stark contrast to the trend observed in the absence of stopping. Conversely, if target parameters fall short, incomplete energy deposition causes the 900 keV protons to yield less fusion than their 672 keV counterparts.

In order to clarify the influence of stopping power on the observed behavior, we performed comparative simulations with $a_0=11.5$, producing a proton beam with a central energy of approximately 900 keV. The $\alpha$-particle
yield is calculated by coupling a Monte Carlo (MC) post-processor\cite{lei2024compact} with PIC simulation. The MC code takes the ion positions, momenta, and densities from the PIC results and spatially bins the macro-particles into individual cells. We can calculate the $\alpha$ production rate $dY_\alpha/dt$ per time step $dt$ through particle pairs:
\begin{equation}
	\frac{dY_{\alpha}}{dt} =  \sum\limits_{i \in \mathcal{C}}  3 n_{p,i} \, n_{B,i} \langle \sigma v \rangle_i
\end{equation}
with
\begin{equation}
	\langle \sigma v \rangle_i =   \frac{ \sum\limits_{j \in \mathcal{P}_i} \sum\limits_{k \in \mathcal{B}_i} w_{pi,j} \, w_{Bi,k} \, \sigma(v_{rel}) v_{rel}}{\sum\limits_{j \in \mathcal{P}_i} \sum\limits_{k \in \mathcal{B}_i} w_{pi,j} \, w_{Bi,k}},
\end{equation}
where $\mathcal{C}$ denotes the set of all grid cells, and $\mathcal{P}_i$ ($\mathcal{B}_i$) represents the set of proton (boron) macroparticles within the $i$-th cell. 
$w_{pi,j}$ and $w_{Bi,k}$ are the statistical weights of the $j$-th proton and $k$-th boron macroparticle, respectively. The number densities of protons and boron in the $i$-th cell are given by $n_{p,i}$ and $n_{B,i}$. $v_{rel} = |\mathbf{v}_{pi,j} - \mathbf{v}_{Bi,k}|$ is the relative velocity between the colliding pair. 
Based on Eq. (6), by integrating the production rate at each time step, we obtain the temporal evolution of the total $\alpha$-particle yield $Y_\alpha =\int (\Delta Y_\alpha/\Delta t) dt $. To facilitate a direct comparison, both the cumulative yield and the production rate are normalized in this section. The fusion yield $Y_\alpha$ is normalized to $1.1 \times 10^{20} m^{-2}$, the rate $\Delta Y_\alpha/\Delta t$ is normalized to $8.3 \times 10^3 m^{-2}$ and $\Delta t$ is chosen to be $5T_0$.

The yield comparison between Figs. 5(a) and 5(b) clearly indicates that stopping power effect significantly boosts the fusion yield. The enhancement originates from the fact that protons, as they decelerate, continuously cover the energy range around the cross-section peak, effectively increasing the probability of fusion reactions.  The theoretical curve in Fig. 5(c), obtained with identical parameters, further corroborates this phenomenon. 

\begin{figure}[t]
	\includegraphics[width=0.48\textwidth]{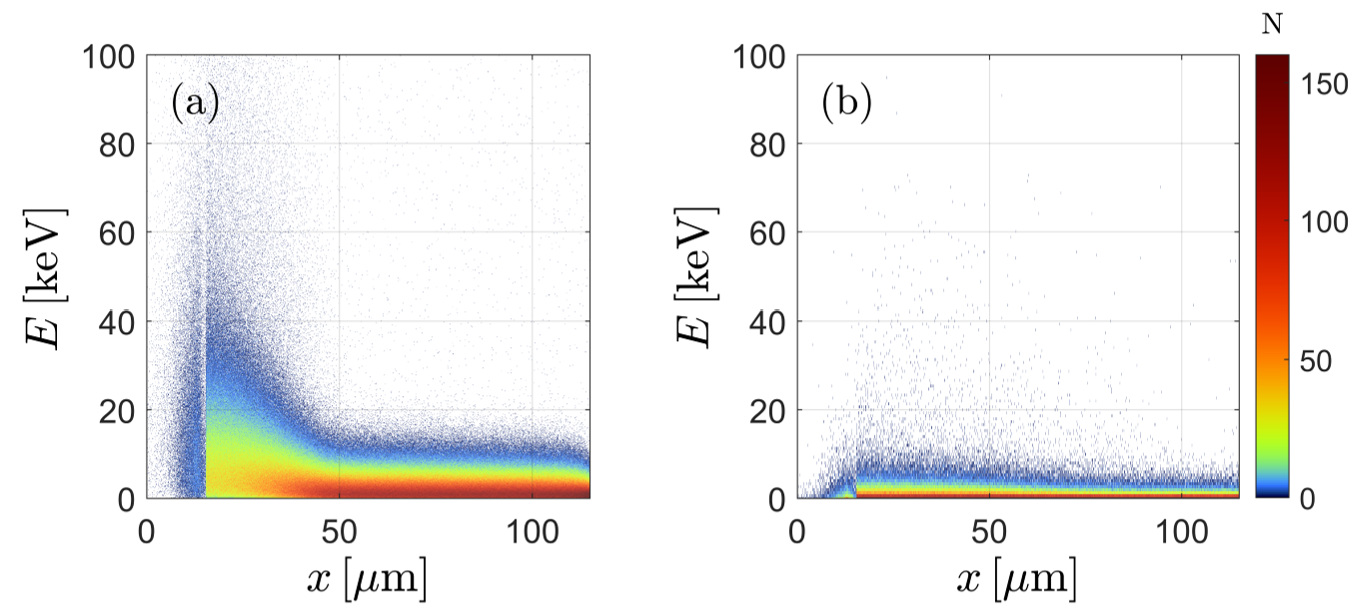}
	\caption{The electron energy-position distribution at $t = 1000T_0$ for the 900 keV proton beam simulation  (a) including stopping power, and (b) excluding stopping power.}\label{fig6}
\end{figure}

\begin{figure}[b]
	\includegraphics[width=0.32\textwidth]{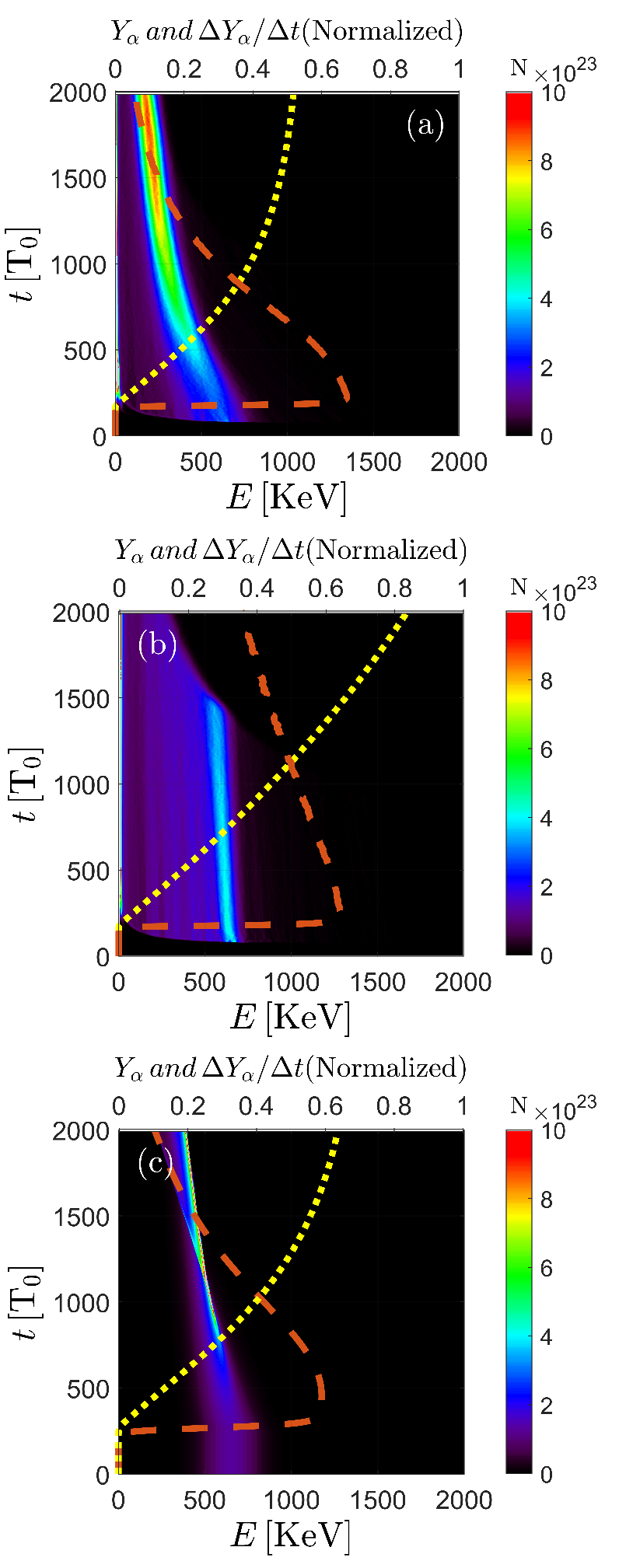}
	\caption{Comparative analysis of the $\alpha$-particle yield generated by a 672 keV proton beam interacting with a boron target, where the laser intensity $a_0=9.5$. The other parameters are same with them in Fig. 5.}
	\label{fig7}
\end{figure}

The reaction rate reaches its maximum at approximately $1000 T_0$ (orange dashed line) when the central energy of the proton beam approaches 672 keV. As the proton beam propagates through the plasma, its central energy decreases while the relative energy spread increases, as shown in Fig. 5(a). This behavior is also well explained by the theoretical model, which was presented in Fig. 5(c). The faster protons at the forefront transfer their kinetic energy to the background electrons via collisional heating, leading to a substantial elevation of the local electron temperature within the target. This localized heating, in turn, dynamically suppresses the effect of stopping power, which preferentially mitigates the deceleration of the trailing protons. Owing to this mechanism, the relative energy spread of the proton beam is effectively maintained at a relatively low level.
 In contrast, this phenomenon disappears in the absence of stopping power, as revealed in Fig. 5(b).

Furthermore, the feedback effect is further substantiated by the electron kinetic energy distribution, as shown in Figs. 6(a) and 6(b). Owing to the stopping power effect, the electron energy within the proton beam propagation region is significantly increased. To further illustrate the feedback effect of stopping power, we use a proton beam with a central energy of 672 keV, corresponding to a laser intensity of $a_0=9.5$. The results of these simulations, including both the cases with and without stopping power, are presented in Figs. 7. The incorporation of stopping power leads to a reduced fusion yield relative to the no‑stopping case, which is consistent with the theoretical predictions in Figs. 4(a) and 4(b).

More importantly, in the presence of stopping power, the 900 keV proton beam produces a greater fusion yield than the 672 keV beam, as presented in Fig. 5(a) and Fig. 7(a). This outcome represents a complete reversal of the trend observed in the absence of stopping power, where the lower‑energy beam was more efficient. On the one hand, the stopping power leads to energy loss of the protons within the target, shifting the relevant portion of the fusion cross-section. On the other hand, beam–target interactions raise the electron temperature, which effectively lowers the average stopping power and enables more efficient proton propagation. This finding indicates that the optimal proton energy for the pitcher-catcher scheme is not strictly fixed at 672 keV, but rather depends on the target conditions. For boron targets of fixed density and spatial scale, the stopping effect, combined with electron-temperature feedback, renders the 900 keV beam more efficient in terms of fusion yield.

\section{Conclusion}

In this work, we develop a  theoretical model for the stopping power of ion beams in hot plasmas, with a specific focus on the feedback mechanism mediated by electron temperature variations. It is found that, when stopping power is included, the energy deposition dynamics play a critical role in determining the effective fusion cross section sampled by the protons, thereby requiring a reassessment of the optimal beam energy. Contrary to the 672 keV resonance peak of the intrinsic cross section, the optimal beam energy is shifted to 900 keV to maximize the fusion yield under the investigated conditions. Furthermore, this phenomenon is validated through PIC simulation. The present work therefore establishes a comprehensive reference for beam target interaction dynamics in fusion relevant plasmas and provides practical guidance for optimizing pitcher–catcher targets in future fusion experiments.

\section{ACKNOWLEGEMENT}

This work was supported by the Strategic Priority Research Program of Chinese Academy of Sciences (No. XDB0890200), the National Natural Science Foundation of China (12388102, 12325409, U2267204, 12405281 and 12474350),  the Natural Science Foundation of Shanghai(24ZR1493000) and the Youth Innovation Promotion Association of Chinese Academy of Sciences.

\section*{AUTHOR DECLARATIONS}

\textbf{CONFLICT OF INTEREST}\\ 
The authors declare no competing financial interests.

\textbf{AUTHOR CONTRIBUTIONS}\\
X. F. Li and Y. Tian proposed the idea, supervised the work and improved the manuscript,  J. Y. Hua developed the theoretical model, carried out all simulations, analysed the results and drafted the manuscript, X. F. Li, J. X. Wang, Y. X. Leng, Y. Tian, and R. X. Li improved the manuscript. All authors discussed the results, commented on the manuscript, and agreed on the contents.

\textbf{DATA AVAILABILITY} 
The data that support the findings of this study are available from the corresponding authors on request.\\

\vspace{2\baselineskip}
\textbf{References}
\bibliography{ref}
\end{document}